\documentstyle[12pt,epsfig]{article}
\def\be{\begin{equation}} \def\ee{\end{equation}} \def\bea{\begin{eqnarray}}
\def\eea{\end{eqnarray}} \def\nnb{\nonumber}

\begin{document}

\hfill{July 25, 2007}

\begin{center}
\vskip 10mm 
{\Large\bf
$pp\to pp\pi^0$ near threshold 
in pionless effective theory 
}
\vskip 8mm 
{\large 
Shung-ichi Ando\footnote{E-mail:sando@meson.skku.ac.kr} } 
\vskip 8mm
{\large \it 
Department of Physics, Sungkyunkwan University,
Suwon 440-746, Korea 
}
\end{center}

\vskip 10mm

The total cross section of the $pp\to pp\pi^0$ reaction
near threshold is calculated in 
pionless effective field theory 
with a di-baryon and external pions. 
The amplitudes for a leading one-body 
and subleading contact neutral pion production vertex
are obtained
including the initial and final state interactions.
After estimating a low-energy constant in the contact vertex, 
we compare our results for the total cross section
with the experimental data. 

\newpage 

\vskip 1cm

\vskip 2mm \noindent 
{\bf 1. Introduction} 

The study of neutral pion production in proton-proton collision
near threshold, $pp\to pp\pi^0$, 
has been inspired by precise measurements of the near-threshold
cross section~\cite{metal-prl90,metal-npa92}.
Surprisingly, the measured cross section 
turned out to be $\sim$ 5 times larger
than the early theoretical predictions~\cite{kr-pr66,ms-prc91}.
Main theoretical difficulties in describing 
the threshold cross section may stem from the suppression mechanisms 
in both one- and two-body $S$-wave neutral pion production operators 
and the typical large momentum transfer $k$ 
between two protons at the threshold, 
$k\simeq \sqrt{m_\pi m_N}$ ($k^{-1}\sim 0.55$ fm), 
where $m_\pi$ and $m_N$ are
the pion and nucleon mass, respectively.
Subsequently, some mechanisms to account for 
the threshold experimental data 
have been suggested in model calculations: 
one is the short-range effect 
of heavy-meson exchanges~\cite{lr-prl93} 
and another is the off-shell effect of 
$\pi N$ $S$-wave isoscalar amplitude
in the one-pion exchange production operator~\cite{ho-plb95}.

Heavy-baryon chiral perturbation theory (HB$\chi$PT)
is a low-energy effective field theory (EFT) of QCD 
and provides us
a systematic perturbation scheme 
in terms of $Q/\Lambda_\chi$ where
$Q$ denotes small external momentum and/or symmetry breaking term 
$m_\pi$ and $\Lambda_\chi$ denotes the chiral scale 
$\Lambda_\chi=4\pi f_\pi\simeq 1$ GeV: 
$f_\pi$ is the pion decay constant.
(For reviews, see, e.g., 
Refs.~\cite{betal-00,bk-arnps02,kp-arnps04,e-ppnp06,b-07}.)
In Refs.~\cite{petal-prc96,cetal-prc96,setal-prc97}
the total cross section of $pp\to pp\pi^0$ was calculated
at the tree-level, 
where the production operators
were derived from the HB$\chi$PT Lagrangian 
up to next-to leading order (NLO)
with the Weinberg's counting rules~\cite{weinberg},
and the matrix element was obtained by 
DWBA: thus the operators were sandwiched
between the initial and final two-nucleon wavefunctions 
obtained by solving the Schr\"{o}dinger equation
with the accurate phenomenological $NN$ potentials.
Up to one-loop order, i.e., 
next-to-next-to leading order (NNLO),
the production operators were derived in Ref.~\cite{detal-plb99} 
and the total cross section was obtained by including the initial
and final state interactions 
in Refs.~\cite{aetal-plb01,ketal-07}. 
The other approaches, e.g.,
a tree-level calculation including heavy-mesons~\cite{vkmr-plb96}
and a relativistic calculation including 
a part of the one-loop diagrams~\cite{bkm-epja99}
were also reported.
For a recent review, see Ref.~\cite{h-pr04} and references therein.

Though many works on the $pp\to pp\pi^0$ reaction near 
threshold in HB$\chi$PT have been done,
some issues in theoretically describing the process 
have not been fully clarified. 
In the NLO HB$\chi$PT calculations~\cite{petal-prc96,cetal-prc96,setal-prc97}, 
a significant enhancement of the off-shell $\pi\pi NN$
vertex function obtained from the NLO HB$\chi$PT Lagrangian 
is found. However,
the two-body (one-pion-exchange) matrix element 
with the off-shell $\pi\pi NN$ vertex 
is almost exactly canceled with the one-body 
matrix element. 
Thus the experimental data cannot be reproduced in the NLO calculations.
In the NNLO HB$\chi$PT 
calculations~\cite{detal-plb99,aetal-plb01,ketal-07},
a significant contribution 
comes out of the NNLO corrections 
and a moderate agreement with the experimental data 
is obtained~\cite{aetal-plb01}. 
However, the chiral series 
based on the standard Weinberg's counting rules
shows poor convergence.
A modification of the original Weinberg's counting rules 
to account for the large momentum transfer, 
$k\simeq \sqrt{m_\pi m_N}$, 
is discussed in Ref.~\cite{cetal-prc96}.
The production operators at NLO using the modified counting rules 
are estimated, and it was reported that 
the NLO contributions exactly cancel
among themselves~\cite{hk-prc02}.
Recently, some detailed issues
for the loop calculations, such as
a concept of reducibility~\cite{letal-epja06},
a representation invariance of the chiral fields 
among the loop diagrams, 
and a proper choice of the heavy-nucleon 
propagator~\cite{hw-plb07}, 
were also studied.\footnote{
Some other issues, 
a role of three-particle singularities
for the $\pi NN$ system (using a toy model)~\cite{hetal-prc01}
and the use of the effective low-momentum $NN$ potential 
$V_{low\mbox{-}k}$ for the process~\cite{ketal-prc06},
were also studied.
}

The problem in 
those 
NNLO HB$\chi$PT calculations 
for the process is that 
no hierarchy in constructing the production operators
is found.
A plausible explanation for this situation is that 
along with the suppression of the leading order Weinberg-Tomozawa
term in the $\pi\pi NN$ vertex for the one-pion exchange contribution,
the one pion exchange propagator, which is usually counted as $Q^{-2}$,
gains another suppression factor $Q/m_N$ because of the typical 
large momentum transfer $k^2\simeq m_\pi m_N$ and thus is counted
as $Q^{-1}\sim (m_\pi m_N)^{-1}$.
Therefore the one-pion-exchange diagram for the process 
could gain the two suppression 
factors $(Q/\Lambda_\chi)^2$ and thus become the same order as the 
one-loop diagrams in the original Weinberg's counting rules.
Apparently the contributions of
the intermediate two-pion exchange  and 
the short-range contact diagrams
play almost the same role as
that of the long-range one-pion-exchange one.
A possible way to improve the situation 
would be to calculate higher order corrections
employing the modified counting rules
or to employ a relativistic formalism 
which we will discuss later.
Because performing higher order loop corrections 
is a formidable task, 
an easy way to circumvent the problem
might be to employ a pionless theory in which 
virtual pions exchanged between the two nucleons
are integrated out;
in this pionless theory, the one-pion exchange, 
two-pion exchange and contact terms in HB$\chi$PT
are subsumed in a contact term. 

In this work we employ a pionless effective field theory (EFT) 
with a di-baryon~\cite{bs-npa01,ah-prc05}\footnote{
We have studied $np\to d\gamma$ cross section 
at BBN energies~\cite{achh-prc06} and 
neutron-neutron fusion process~\cite{ak-plb06} 
employing this formalism.} 
and external pions~\cite{bs-npa03}
to calculate the total cross section of 
the $pp\to pp\pi^0$ process. 
A main motivation of this work stems from the observation
that some of the HB$\chi$PT calculations 
failed in reproducing the experimental data.
Meanwhile, it is known that the energy dependence of the 
experimental data can be well reproduced in terms of
the final state interaction and the phase space~\cite{metal-npa92}.
The pionless theory would be a ``minimal'' formalism to
take account of these two features.
Furthermore, after taking these two features into account,
the difference between the theory and experiment
appears in the overall factor and 
the experimental data can be easily reproduced
by fitting an unknown constant
that appears in a contact vertex.

This paper is organized as follows.
In Sec.~2 the pionless effective Lagrangian with
a di-baryon and external pions is introduced. 
In Sec.~3 we fix low-energy constants (LECs) 
for the final $^1S_0$  
and initial $^3P_0$ $NN$ states.
In Sec.~4 the amplitudes of $pp\to pp\pi^0$ 
for a leading one-body and subleading contact vertex
are obtained including the strong initial state interaction 
and the strong and Coulomb final state interactions.
We estimate in Sec.~5 the value of 
an LEC in the external neutral pion production contact vertex
from HB$\chi$PT, 
and in Sec.~6 we show
our numerical results of the total cross section
and compare with the experimental data.
Finally, in Sec.~7, the discussion and conclusions of this work
are given.

\vskip 5mm \noindent 
{\bf 2. Pionless effective Lagrangian 
with a di-baryon and external pions} 

An effective Lagrangian without virtual pions 
and with a di-baryon and external pions 
for describing the $pp\to pp\pi^0$ reaction may read 
\bea
{\cal L} = 
  {\cal L}_N   
+ {\cal L}_s 
+ {\cal L}_{Ns\pi} 
+ {\cal L}_{NN}^P \, ,
\eea
where ${\cal L}_N$ is the one-nucleon Lagrangian
interacting with the external pions, 
${\cal L}_{s}$ is that for the 
$^1S_0$ channel   di-baryon field
interacting with two-nucleon,
${\cal L}_{Ns\pi}$ is 
the contact interaction Lagrangian for external pion-dibaryon-two-nucleon, 
and ${\cal L}_{NN}^P$ is the two-nucleon Lagrangian 
for the $^3P_0$ channel.

The one-nucleon Lagrangian ${\cal L}_N$ in heavy-baryon
formalism reads
\bea
{\cal L}_N =
N^\dagger 
\left\{iv\cdot D + 2i g_A S\cdot \Delta 
\frac{}{} 
+ \frac{1}{2m_N} \left[
(v\cdot D)^2-D^2 
+2g_A\{v\cdot \Delta,S\cdot D\}
+\cdots 
\right]
\right\} N,
\label{eq;L1}
\eea
where $v^\mu$ is a velocity vector with a condition $v^2=1$
where $v^\mu=(1,\vec{0})$, 
and the spin operator $S^\mu$ is $S^\mu= (0,\vec{\sigma}/2)$.
$D_\mu = \partial_\mu + \Gamma_\mu$ with
$\Gamma_\mu= \frac12[\xi^\dagger,\partial_\mu \xi]$ and 
$\Delta_\mu= \frac12\{\xi^\dagger,\partial_\mu\xi\}$. 
The nonlinearly realized external pions 
in the $\xi$ field are given by 
$\xi = {\rm exp}[i\vec{\tau}\cdot \vec{\pi}/(2f_\pi)]$.
$g_A$ is the axial vector coupling. 

The effective Lagrangian for the two-nucleon part
may read~\cite{bs-npa01,ah-prc05,crs-npa99,fms-npa00}
\bea
{\cal L}_s &=&
\sigma_s s_a^\dagger\left[iv\cdot D
+\frac{1}{4m_N}[(v\cdot D)^2-D^2]
+\delta_s\right] s_a
-y_s\left[s_a^\dagger (N^TP^{(^1S_0)}_aN) + \mbox{h.c.}\right], 
\\
{\cal L}_{Ns\pi} &=& 
\frac{\tilde{d}_\pi^{(2)}}{\sqrt{8 m_Nr_0}}
\left\{
i\epsilon_{abc} 
s_a^\dagger 
\left[
N^T\sigma_2 \vec{\sigma}\cdot i 
(\stackrel{\to}{D}
-\stackrel{\leftarrow}{D})
\tau_2\tau_b N
\right] (iv\cdot \Delta_c)
+ \mbox{\rm h.c.}
\right\}
\, ,
\\
{\cal L}_{NN}^P &=&  
C_2^0 \delta_{ij}\delta_{kl}\frac14
\left(N^T{\cal O}_{ij,a}^{1,P}N\right)^\dagger
\left(N^T{\cal O}_{kl,a}^{1,P}N\right) 
+ \cdots \, ,
\eea
with 
\bea
{\cal O}_{ij,a}^{1,P} =
i(\stackrel{\leftarrow}{D}_i P_{j,a}^{(P)} 
-P_{j,a}^{(P)} \stackrel{\rightarrow}{D}_i)
\, ,
\ \ \ 
P_{i,a}^{(P)} = \frac{1}{\sqrt{8}}\sigma_2 \sigma_i \tau_2\tau_a\, ,
\eea
where 
$s_a$ is the (spin singlet) $^1S_0$ channel di-baryon field
with the isospin index $a$, 
$\sigma_{s}$ is the sign factor $\sigma_{s}=\pm 1$.
$\delta_s$ is the mass difference
between the di-baryon mass $m_s$ 
and two-nucleon mass, i.e.,
$m_{s} = 2 m_N + \delta_{s}$.
$y_{s}$ is the coupling constant
of the di-baryon and two-nucleon interaction.
$P_i^{({\cal S})}$ are the projection operators 
for ${\cal S}={}^1S_0$ and $^3P_0$ channels;
\bea
P_a^{({}^1S_0)} = \frac{1}{\sqrt{8}}\sigma_2\tau_2\tau_a\, ,
\ \ 
P^{(^3P_0)}_a = \frac{1}{\sqrt{8}}
\sigma_2\vec{\sigma}\cdot\hat{p}\tau_2\tau_a\, ,
\int\frac{d\Omega_{\hat{p}}}{4\pi}\sum_{pol. avg}
{\rm Tr}\left(P^{({\cal S})\dagger}
P^{({\cal S}')}\right) 
= \frac12\delta^{{\cal SS}'}\, , 
\eea 
where $\sigma_i$ ($\tau_a$) is the spin (isospin) operator.
$d_\pi^{(2)}$ is an unknown LEC of the 
(external) pion-(spin singlet) dibaryon-nucleon-nucleon ($\pi sNN$) 
interaction.
$r_0$ is the effective range in the $^1S_0$ ($pp$) channel,
and $\Delta^\mu=\frac{\tau_a}{2}\Delta^\mu_a$. 
$C_2^0$ is the LEC for the $P$-wave $NN$ scattering in the $^3P_0$ 
channel.

\vskip 2mm \noindent
{\bf 3. Fixing LECs of the initial and final $NN$ interactions}

\begin{figure}
\begin{center}
\epsfig{file=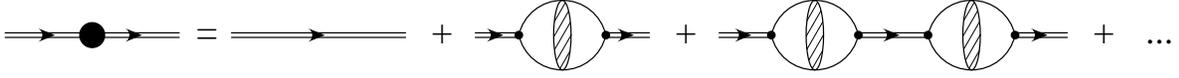,width=15.5cm}
\caption{
Diagrams for the dressed di-baryon propagator 
including the Coulomb interaction.
A double-line with a filled circle denotes 
the renormalized dressed di-baryon propagator.
Double-lines without a filled circle and 
single-line-curves 
denote the bare di-baryon propagators and  
nucleon propagators, respectively.
The two-nucleon propagator with
a shaded blob denotes the Green's function 
including the Coulomb potential.
A (spin-singlet) dibaryon-nucleon-nucleon ($sNN$) vertex is proportional
to the LEC $y_s$.
\label{fig;dibaryon-propagator}}
\end{center}
\end{figure}

In this section, we calculate the $S$- and $P$-wave $NN$ 
scattering amplitudes to fix the LECs in the two-nucleon part.
In Fig.~\ref{fig;dibaryon-propagator},
diagrams for the dressed $^1S_0$ channel di-baryon propagator 
are shown where the two-nucleon bubble diagrams 
including the Coulomb interaction
are summed up to the infinite order.
The inverse of the propagator 
in the center of mass (CM) frame is given  by
\bea
i D_{s}^{-1}(p) &=& 
i\sigma_{s}(E+\delta_{s})-iy_{s}^2J_0(p)  \, ,
\eea
with
\bea
J_0(p) &=& 
\int\frac{d^3\vec{k}}{(2\pi)^3}
\frac{d^3\vec{q}}{(2\pi)^3}
\langle \vec{q}| \hat{G}_C^{(+)}(E)|\vec{k}\rangle\, ,
\eea 
where $\hat{G}_C^{(+)}$ is the outgoing two-nucleon
Green's function including the Coulomb potential,
\bea
\hat{G}_C^{(+)}(E)=\frac{1}{E-\hat{H}_0-\hat{V}_C+i\epsilon}\, ,
\eea
where $E$ is the total CM energy, $E=p^2/m_N+\cdots$,
$\hat{H}_0$ is the free Hamiltonian for two-proton,
$\hat{H}_0=\hat{p}^2/m_N$, and $\hat{V}_C$ is the repulsive 
Coulomb force $\hat{V}_C= \alpha/r$: $\alpha$ is the
fine structure constant. 
Employing the dimensional regularization
in $d=4-2\epsilon$ space-time dimensions,
we obtain~\cite{kr-plb99,ashh-07} 
\bea
J_0(p) &=& 
\frac{\alpha m_N^2}{8\pi}\left[
\frac{1}{\epsilon}
-3\gamma
+ 2
+{\rm ln}\left(\frac{\pi\mu^2}{\alpha^2m_N^2}\right)
\right]
-\frac{\alpha m_N^2}{4\pi}h(\eta)
-C_\eta^2\frac{m_N}{4\pi}(ip)\, ,
\eea
where $\mu$ is the scale of the dimensional regularization,
$\gamma= 0.5772\cdots$,
and 
\bea
&& h(\eta) = Re\, \psi(i\eta)-{\rm ln}\eta\, ,
\ \ \ 
Re\, \psi(\eta) = 
\eta^2 \sum_{\nu=1}^\infty \frac{1}{\nu(\nu^2+\eta^2)}
-\gamma\, ,
\nnb \\ &&
C_\eta^2 = \frac{2\pi \eta}{e^{2\pi\eta}-1}\, ,
\ \ \ 
\eta= \frac{\alpha m_N}{2p}\, .
\eea
Thus the inverse of the renormalized dressed di-baryon propagator 
is obtained as
\bea
i D_{s}^{-1}(p) &=& 
iy_s^2 \frac{m_N}{4\pi}\left[
\frac{4\pi \sigma_s \delta_s^R}{m_Ny_s^2}
+ \frac{4\pi\sigma_s}{m_N^2y_s^2}p^{2}
+\alpha m_N h(\eta)
+ ip\, C_\eta^2 
\right]\, ,
\eea
where $\delta_s^R$ is the renormalized 
mass difference between the di-baryon and two nucleons, 
\bea
\sigma_s \delta_s^R &=& 
\sigma_s \delta_s
-y_s^2 
\frac{\alpha m_N^2}{8\pi}\left[
\frac{1}{\epsilon}
-3\gamma
+ 2
+{\rm ln}\left(\frac{\pi\mu^2}{\alpha^2m_N^2}\right)
\right] \, .
\eea
We fix it by using the scattering length $a_C$ below.

\begin{figure}[t]
\begin{center}
\epsfig{file=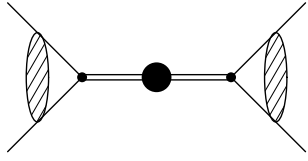,width=4cm}
\caption{
Diagram for the $S$-wave $pp$ scattering amplitude
with the Coulomb interaction.
See the caption of Fig.~\ref{fig;dibaryon-propagator} 
for details.
\label{fig;NNamplitudes}}
\end{center}
\end{figure}
In Fig.~\ref{fig;NNamplitudes}, 
a diagram of the $S$-wave $pp$ scattering 
amplitude with the Coulomb interaction is shown
and 
thus we have the $S$-wave scattering amplitude as
\bea
i{\cal A}_{s} &=& 
(-iy_{s}\psi_0)iD_{s}(p)(-iy_{s}\psi_0)
\nnb \\
&=& 
i \frac{4\pi}{m_N}\frac{C_\eta^2e^{2i\sigma_0}}{
-\frac{4\pi\sigma_{s}\delta_{s}^R}{m_Ny_{s}^2}
-\frac{4\pi\sigma_{s} p^2}{m_N^2y_{s}^2}
-\alpha m_N h(\eta)
-ip\, C_\eta^2} \, , 
\eea
with
\bea
\psi_0 &=& 
\int\frac{d^3\vec{k}}{(2\pi)^3} 
\langle \vec{k}|\psi_{\vec{p}}^{(+)}\rangle 
=\int\frac{d^3\vec{k}}{(2\pi)^3} 
\langle\psi_{\vec{p}}^{(-)}|\vec{k}\rangle 
= C_\eta e^{i\sigma_0}\, ,
\eea
where $\langle\vec{k}|\psi_{\vec{p}}^{(\pm)}\rangle$ are 
the Coulomb wave functions 
obtained by solving the Schr\"{o}dinger equation
$(\hat{H}-E)|\psi^{(\pm)}_{\vec{p}}\rangle = 0$
with $\hat{H}=\hat{H}_0+\hat{V}_C$
and represented in the $|\vec{k}\rangle$ space 
for the two protons.
$\sigma_0$ is the $S$-wave Coulomb phase shift
$\sigma_0={\rm arg}\,\Gamma(1+i\eta)$. 
The $S$-wave amplitude ${\cal A}_s$ is 
given in terms of the effective range parameters 
as 
\bea
i{\cal A}_s &=& 
i \frac{4\pi}{m_N}\frac{C_\eta^2e^{2i\sigma_0}}{
-\frac{1}{a_C} + \frac12 r_0 p^2 + \cdots
-\alpha m_N h(\eta)
-ip\, C_\eta^2} \, ,
\eea
where $a_C$ is the scattering length,
$a_C=-7.8063\pm 0.0026$ fm, 
$r_0$ is the effective range, 
$r_0=2.794\pm 0.014$ fm,
and the ellipsis represents 
the higher order effective range corrections.
Now it is easy to match the LECs 
with the effective range parameters.
Thus we have $\sigma_{s} = -1$ and
\bea
&& y_s = \pm \frac{2}{m_N}\sqrt{\frac{2\pi}{r_0}}\, ,
\ \ \ 
D_s(p) = \frac{m_Nr_0}{2}\frac{1}{\frac{1}{a_C} 
-\frac12r_0 p^2 +\alpha m_N h(\eta) +ip\, C_\eta^2}\, . 
\label{eq;ys}
\eea
We note that the sign of the LEC $y_s$ cannot be determined
by the effective range parameters.

\begin{figure}[t]
\begin{center}
\epsfig{file=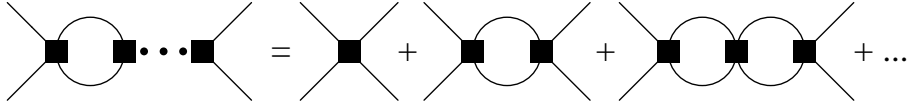,width=12.cm}
\caption{
Diagrams for the $P$-wave $NN$ scattering.
Two-nucleon bubble diagrams
are summed up to the infinite order. 
Lines and curves denote nucleon propagators.
A four-nucleon contact vertex with a filled box
is proportional to the LEC $C_2^0$.
\label{fig;pwavescattering}}
\end{center}
\end{figure}

In Fig.~\ref{fig;pwavescattering},
diagrams for the $P$-wave $NN$ scattering 
are shown. 
Because the momenta of the two protons are quite large 
for the pion production reaction, 
we cannot treat the $P$-wave vertex correction 
in a perturbative way. Thus
the two-proton bubble diagrams 
are summed up to the infinite order 
without including the Coulomb interaction~\footnote{
An effective Coulomb interaction is denoted 
by $\eta = \alpha M/(2p)$. 
For the high momenta $p\ge\alpha M/2\simeq 
3.42$ MeV, $\eta\le 1$ and the Coulomb interaction 
can be treated in a perturbative way.
\label{footnote;Coulomb} }
and the LEC $C_2^0$ is renormalized by a phase shift
at the threshold energy.
The scattering amplitude for 
the ${}^3P_0$ channel is obtained as 
\bea
i{\cal A}_p &=& \frac{4\pi}{m_N}\frac{ip^2}
 {\frac{4\pi}{m_NC_2^0}-ip^3} \, ,
\eea
where we have used the nucleon propagator, 
$iS_N(k) = i/[k_0 - \vec{k}^2/(2m_N)+i\epsilon]$, 
where $k^\mu$ is the residual nucleon momentum 
$k^\mu = P^\mu-m_Nv^\mu$.
$P^\mu$ is the nucleon momentum $P^2=m_N^2$, 
and we have employed the dimensional regularization 
for the loop calculation.
The LEC $C_2^0$ is 
fixed by the phase shift of ${}^3P_0$ channel at 
pion production threshold,
$\delta_p(p_{th})\simeq -7.5^\circ$ at 
$p_{th}\simeq \sqrt{m_\pi m_N}$.
Thus we have
\bea
\frac{4\pi}{m_NC_2^0} \simeq p_{th}^3\cot\delta_p(p_{th}) \, .
\eea

\vskip 2mm \noindent
{\bf 4. Amplitudes for $pp\to pp\pi^0$ near threshold}

\begin{figure}[t]
\begin{center}
\epsfig{file=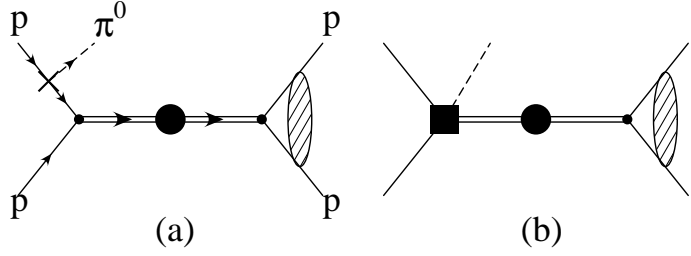,width=9.cm}
\caption{
Diagrams for $pp\to pp\pi^0$  near threshold
with the strong and Coulomb final state interactions
and without the initial state interaction.
$S$-wave neutral pion is emitted 
from $\pi NN$ vertex with ``X'' at the $1/m_N$ order in (a), 
whereas the pion is emitted 
from a pion-dibaryon-nucleon-nucleon contact vertex
in (b) which is proportional to LEC $\tilde{d}_\pi^{(2)}$.
\label{fig;diagrams-non-FSI}}
\end{center}
\end{figure}
\begin{figure}[t]
\begin{center}
\epsfig{file=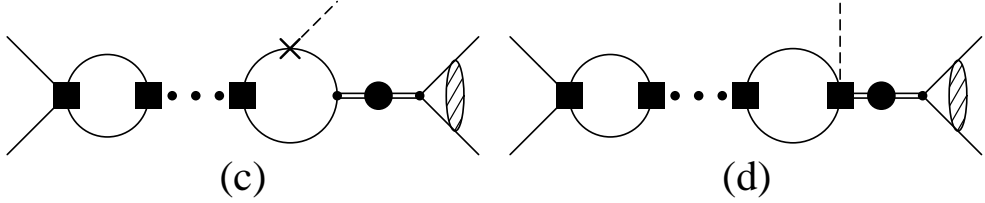,width=13.cm}
\caption{
Diagrams for $pp\to pp\pi^0$ 
with the strong initial and the strong and Coulomb final state interactions.
See the caption of Fig.~\ref{fig;diagrams-non-FSI} for more
details.
\label{fig;diagrams-FSI}}
\end{center}
\end{figure}
In Figs.~\ref{fig;diagrams-non-FSI}
and \ref{fig;diagrams-FSI},
we show diagrams for $pp\to pp\pi^0$ 
near threshold.
In diagram (a) in Fig.~\ref{fig;diagrams-non-FSI} 
and (c) in Fig.~\ref{fig;diagrams-FSI},
the pion is emitted from the one-body $\pi NN$ vertex.
Because the $S$-wave pion production is considered, 
the $\pi NN$ vertex (the vertex with ``X'' in the figures) 
is obtained from the $1/m_N$ Lagrangian. 
In the diagram (b) in Fig.~\ref{fig;diagrams-non-FSI}
and (d) in Fig.~\ref{fig;diagrams-FSI},
the pion is emitted from the pion-(spin singlet) dibaryon-nucleon-nucleon 
($\pi sNN$) contact vertex which is proportional to the unknown LEC 
$\tilde{d}_\pi^{(2)}$. 
The strong and Coulomb final state interactions are considered
in all of the diagrams (a), (b), (c) and (d) 
in Figs.~\ref{fig;diagrams-non-FSI} and \ref{fig;diagrams-FSI},
whereas 
the strong initial state interaction is considered~\footnote{
See the footnote~\ref{footnote;Coulomb}.
}
in the diagrams (c) and (d) in Fig.~\ref{fig;diagrams-FSI}.

The one-body amplitude from the (a) and (c) diagrams
and the two-body (contact) amplitude from the (b) and (d) diagrams 
are obtained as 
\bea
i{\cal A}_{(a+c)} &=& -\frac{4\pi g_A}{m_N^2f_\pi}
\frac{C_{\eta'} e^{i\sigma_0}p}
{\frac{1}{a_C}-\frac12r_0p'^2
+\alpha m_N h(\eta') +ip'\, C_{\eta'}^2} 
\frac{1}{1-\frac{m_NC_2^0}{4\pi}ip^3}\,,
\label{eq;Aac}
\\
i{\cal A}_{(b+d)} &=& 
4\sqrt{\frac{2\pi}{m_N}}\frac{\tilde{d}_\pi^{(2)}}{f_\pi}
\frac{C_{\eta'}e^{i\sigma_0}\omega_q p}
{\frac{1}{a_C} -\frac12 r_0 p'^2 
+\alpha m_N h(\eta') +ip'\, C_{\eta'}^2}
\frac{1}{1-\frac{m_NC_2^0}{4\pi}ip^3}\,,
\label{eq;Abd}
\eea 
where we have used the nucleon propagator,
$iS_N(k) =i/[k_0-\vec{k}^2/(2m_N)+i\epsilon]$ 
in the calculation of the (a), (c), (d) diagrams.
$2\vec{p}$ and $2\vec{p}'$ are the relative three 
momenta between incoming and outgoing two protons, respectively;
$p=|\vec{p}|$ and $p'=|\vec{p}'|$.
$\eta'=\alpha m_N/(2p')$ and 
$\omega_q$ is the energy of outgoing pion,
$\omega_q=\sqrt{\vec{q}^2+m_\pi^2}$:
$\vec{q}$ is the outgoing pion momentum. 
In the loop calculations for the diagrams (c) and (d) 
in Fig.~\ref{fig;diagrams-FSI},
we have employed the dimensional regularization, 
and we have neglected terms involving 
$\vec{q}$ and $p'$, which are considered to be 
small as compared to $p\simeq \sqrt{m_\pi m_N}$. 
We note that there remain no unknown parameters
in the amplitudes except for 
the LEC $\tilde{d}_\pi^{(2)}$
in the two-body (contact) amplitude in Eq.~(\ref{eq;Abd}). 

\vskip 2mm \noindent
{\bf 5.  Estimate of the LEC $\tilde{d}_\pi^{(2)}$ from HB$\chi$PT}

\begin{figure}[t]
\begin{center}
\epsfig{file=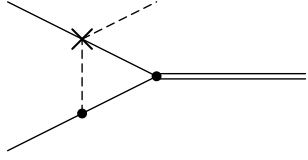,width=4.cm}
\caption{
Diagram for a one-pion exchange contribution
to the $pp\to pp\pi^0$ process for estimation 
the LEC $\tilde{d}_\pi^{(2)}$ in the contact vertex. 
A double-line, single-line, and dashed-line 
denote a di-baryon, nucleon, and pion, respectively.
A $\pi NN$ vertex with a dot is obtained from ${\cal L}_{\pi N}^{(1)}$
and a $\pi\pi NN$ vertex with ``X'' is from
${\cal L}_{\pi N}^{(2)}$ and ${\cal L}_{\pi N}^{(3)}$.
\label{fig;diagrams-loops}}
\end{center}
\end{figure}

In this section we 
estimate an order of magnitude of the LEC $\tilde{d}_\pi^{(2)}$ 
from HB$\chi$PT.  
We here consider 
a one-pion-exchange (OPE) 
diagram shown in Fig.~\ref{fig;diagrams-loops}.
This diagram is the lowest order OPE contribution 
in the standard Weinberg counting rules. 
We include a higher order (relativistic) correction 
to the $\pi\pi NN$ vertex which is found to 
be important~\cite{bkm-epja99} and is,
in the modified counting rules, 
of the same order
as the lowest order diagram. 
Because other diagrams,
e.g, two-pion exchange diagrams
would give comparable contributions to that of 
the LEC $\tilde{d}_\pi^{(2)}$
(see Refs.~\cite{aetal-plb01,ketal-07}), 
we expect that the estimate of the diagram 
in Fig.~\ref{fig;diagrams-loops} would be reliable 
only for the order of magnitude estimation for 
the LEC $\tilde{d}_\pi^{(2)}$.

The effective chiral Lagrangian relevant to this purpose
reads
\bea
{\cal L} = {\cal L}_\pi + {\cal L}_{\pi N}  \, ,
\eea
where ${\cal L}_\pi$ is the $\chi$PT Lagrangian
for the pions, and ${\cal L}_{\pi N}$ is
the HB$\chi$PT Lagrangian 
for the pions and nucleon.
These Lagrangians are expanded as 
\bea
{\cal L}_\pi =  {\cal L}_\pi^{(2)} + \cdots\, ,
\ \ \
{\cal L}_{\pi N} =  {\cal L}_{\pi N}^{(1)} + {\cal L}_{\pi N}^{(2)} 
+ {\cal L}_{\pi N}^{(3)}
+ \cdots\, ,
\eea
where ${\cal L}_\pi^{(2)}$ is the standard chiral Lagrangian
for pions at LO.
The expression of ${\cal L}_{\pi N}^{(1)}$ and some part of 
${\cal L}_{\pi N}^{(2)}$ 
are given in Eq.~(\ref{eq;L1}).
To calculate the isoscalar $\pi\pi NN$ interaction 
in the diagram in Fig.~\ref{fig;diagrams-loops}, 
we consider the interaction Lagrangian~\cite{fmms-ap00}
\bea
{\cal L}_{\pi N}^{(2)} &=& N^\dagger 
\left[
c_1 {\rm Tr}(\chi_+)
+ \left(
\frac{g_A^2}{2m_N}-4c_2\right)
(v\cdot \Delta)^2
-4c_3\Delta\cdot \Delta
\right] N +\cdots\, .
\label{eq;L2}
\\
{\cal L}_{\pi N}^{(3)} &=& 
N^\dagger \left[
\frac{g_A^2}{4m_N^2}\left(
iv\cdot \Delta \Delta \cdot D + h.c. 
\right)
-\frac{2c_2}{m_N}[ 
i {\rm Tr}(v\cdot \Delta \Delta_\mu) D^\mu 
+ h.c.] 
\right] N + \cdots\, ,
\label{eq;L3}
\eea 
where we have included relativistic corrections in
the higher order Lagrangian ${\cal L}_{\pi N}^{(3)}$ 
as mentioned above.
The values of the LECs $c_1$, $c_2$ and $c_3$ 
are fixed in the tree-level calculations~\cite{bkm-npb95}~\footnote{
Values of the LECs $c_1$, $c_2$, and $c_3$ fixed in the one-loop 
calculations are quite different from those in the tree-level ones.
See, e.g., Refs.~\cite{b-07,m-pos05} for details.
} as
\bea
c_1 = -0.64\, , 
\ \ \
c_2 = 1.79 \, , 
\ \ \
c_3 = -3.90\ 
\ [\mbox{\rm GeV$^{-1}$}]\, .
\label{eq;c123}
\eea

Thus a contribution to the $\pi sNN$ vertex function
from the diagram in Fig.~\ref{fig;diagrams-loops}
is obtained as
\bea
\lefteqn{i\Gamma_{\pi sNN} =
-i \frac{m_Ny_sg_A|\vec{p}|\epsilon_{abc}}{16\pi f_\pi^3}
\left[
\left(
-4c_1 +2 c_2 -\frac{3g_A^2}{16m_N} +c_3
\right) m_\pi^2
\int^1_0dx\frac{1-x}{\sqrt{F}}
\right.}
\nnb \\ && \left.
-\left(c_2-\frac{g_A^2}{16m_N}\right)m_\pi^2
\int^1_0dx \frac{x^2(1-x)}{\sqrt{F}}
+\left(c_2-\frac{g_A^2}{16m_N}\right)\frac{m_\pi}{m_N}
\int^1_0dx
(3-5x)\sqrt{F}
\right]\, ,
\eea
with
\bea
F= x(1-x)\vec{p}^2 + x\, m_\pi^2\, ,
\eea
where $a$ and $b$ (in $\epsilon_{abc}$) 
are the isospin indices for the initial and 
final two-nucleon state, respectively, 
and $c$ is that for the outgoing pion.
In the calculation of the diagram 
in Fig.~\ref{fig;diagrams-loops},
we have used the nucleon propagator, 
$iS_N(k) = i/[k_0-\vec{k}^2/(2m_N)+i\epsilon]$,
and the ``potential'' pion propagator,
$i\tilde{\Delta}_\pi(k)=-i/(\vec{k}^2+m_\pi^2-i\epsilon)$
because the typical momentum and energy transfer between two protons
are $\vec{k}^2 \simeq m_\pi m_N$ and $k_0\simeq m_\pi/2$, respectively. 
Furthermore because $\vec{p}^2\simeq m_\pi m_N >> m_\pi^2$,
we take an approximation $F\simeq x(1-x)\vec{p}^2$ and 
thus have
\bea
i\Gamma_{\pi sNN} &\simeq& 
\mp i\frac{\sqrt{2\pi}g_A}{16f_\pi\sqrt{r_0}}
\frac{m_\pi^2}{f_\pi^2}\left(
-4c_1+2c_2
-\frac{3g_A^2}{16m_N} +c_3
\right) \epsilon_{abc} \, .
\eea
Therefore the value of the LEC $\tilde{d}_\pi^{(2)}$ 
from the loop diagram 
in Fig.~\ref{fig;diagrams-loops}
is obtained as
\bea
\tilde{d}_\pi^{(2)}\simeq \pm \frac{\sqrt{2\pi}g_A}{32m_\pi^{3/2}}
\frac{m_\pi^2}{f_\pi^2}\left(
-4c_1
+2c_2
-\frac{3g_A^2}{16m_N}
+c_3
\right)
\simeq \pm 0.140\ \ \mbox{\rm fm$^{5/2}$}\, ,
\label{eq;dpi2}
\eea
where the different signs for  
$\tilde{d}_\pi^{(2)}$
have been obtained because of the LEC $y_s$ 
in Eq.~(\ref{eq;ys}).

\vskip 2mm \noindent
{\bf 6. Numerical results}

Total cross section of $pp\to pp\pi^0$ near threshold 
is calculated using the formula
\bea
\sigma = \frac12 \int^{q^{max}}_0dq \frac{d\sigma}{dq}\, ,
\ \ \ 
\frac{d\sigma}{dq} = \frac{1}{v_{lab}} 
\frac{m_Nq^2p'}{16(2\pi)^3\omega_q}\sum_{spin}|{\cal A}|^2\, ,
\label{eq;sigma}
\eea
with
\bea
p' =|\vec{p}'| \simeq \sqrt{m_N(T-\sqrt{m_\pi^2+q^2})-q^2/4}\, ,
\ \ \
q^{max} \simeq \sqrt{\frac{T^2-m_\pi^2}{1+\frac{T}{2m_N}}}\, ,
\eea
where 
$q$ is the outgoing pion momentum,
$q=|\vec{q}|$, 
$2p$ and $2p'$ are relative momenta of the initial and
final two protons, respectively.  
$T$ is the initial total energy $T\simeq\vec{p}^2/m_N$ and 
$v_{lab}\simeq 2p/m_N$.
We have expanded the proton energies 
in the phase factor
in terms of $1/m_N$ and kept 
up to the $1/m_N$ order. 
${\cal A}$ is the amplitude 
${\cal A} = {\cal A}_{(a+c)} + {\cal A}_{(b+d)}$ 
where ${\cal A}_{(a+c)}$ and ${\cal A}_{(b+d)}$ 
are obtained 
in Eqs.~(\ref{eq;Aac}) and (\ref{eq;Abd}), respectively.
Note that the factor 1/2 in Eq.~(\ref{eq;sigma}) 
is the symmetry factor for 
the final two protons.

\begin{figure}[t]
\begin{center}
\epsfig{file=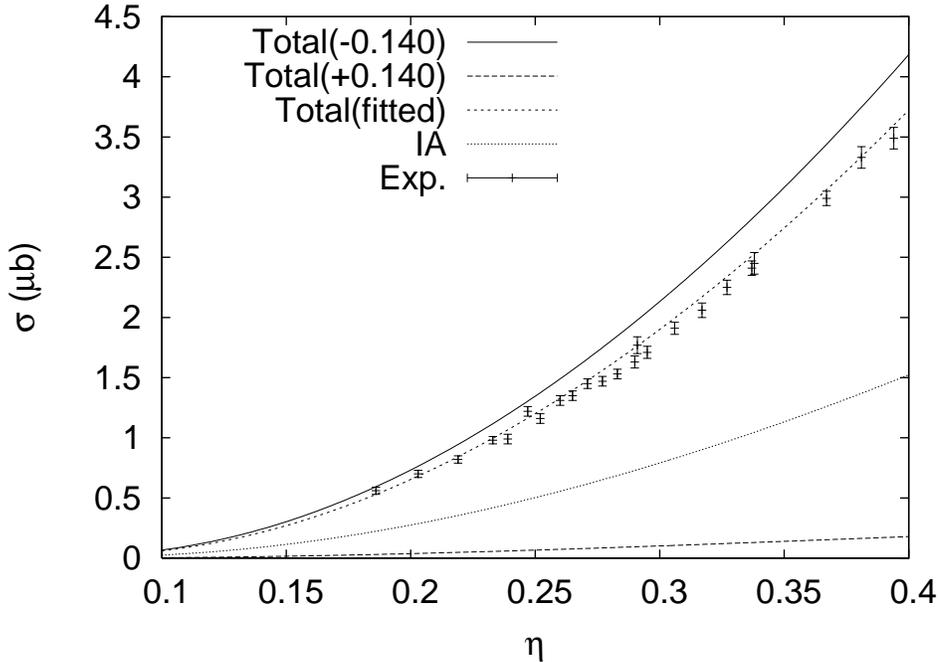,width=13cm}
\caption{
Estimated total cross section of $pp\to pp\pi^0$ 
as a function of $\eta_\pi=|\vec{q}|_{max}/m_\pi$.
See the text for details.
\label{fig;CS}}
\end{center}
\end{figure}

In Fig.~\ref{fig;CS} we plot our results 
of the total cross section
as a function of  
$\eta_\pi= q^{max}/m_\pi$. 
The solid curve and long-dashed curve have been obtained
by using $\tilde{d}_\pi^{(2)}=\pm 0.140$ fm$^{5/2}$ 
fixed from the one-pion exchange diagram 
in Fig.~\ref{fig;diagrams-loops}
in the previous section.
The LEC $\tilde{d}_\pi^{(2)}$ is also fixed by using 
the experimental data as 
\bea
\tilde{d}_\pi^{(2)fitted} = - 0.12 \, , \ \ +0.55 \ \mbox{\rm fm}^{5/2}\, ,
\eea
where we have two values of $\tilde{d}_\pi^{(2)}$ with different signs.
The short-dashed curve 
is obtained by
using $\tilde{d}_\pi^{(2)fitted}= -0.12$ fm$^{5/2}$. 
The dotted line corresponds to the case where
only the contribution from the one-body 
amplitude ${\cal A}_{(a+c)}$ is considered.
The experimental data are also included 
in the figure. 

We find that 
the experimental data are 
reproduced reasonably well
with the value of $\tilde{d}_\pi^{(2)} = -0.14$ fm$^{5/2}$.
By contrast,  we obtain almost vanishing total cross sections
with the value $\tilde{d}_\pi^{(2)}=+0.140$ fm$^{5/2}$
because the two-body amplitude 
with $\tilde{d}_\pi^{(2)}=+0.140$ fm$^{5/2}$
is almost canceled with the amplitude from the one-body 
contribution. 
On the other hand, 
for the whole energy range the experimental 
near-threshold cross section data
are well reproduced with the use of  the fitted parameter 
$\tilde{d}_\pi^{(2)fitted}=-0.12$ fm$^{5/2}$.

We also find that 
approximately a half of the 
experimental data comes from 
the one-body (IA) amplitude 
in the pionless theory.
Whereas, in the previous DWBA calculations,
about 1/5 of the experimental data
comes from the one-body matrix element.
The different one-body contributions in
the pionful and pionless theory stem from 
the different
short range contributions 
in the pionless and pionful theory.
The importance of the short range contributions
in the one-body part of the DWBA calculations 
has been pointed out 
(see, e.g., in Fig.~10 in Ref.~\cite{cetal-prc96}).
In the pionless theory, we conjecture that 
some short range part 
of the one-body matrix element in the DWBA calculations
is integrated out 
and held in the contact LEC $\tilde{d}_\pi^{(2)}$.

\vskip 2mm \noindent
{\bf 7. Discussion and conclusions}

In this work we calculated the total cross section 
for $pp\to pp\pi^0$ near threshold
in pionless EFT with the di-baryon and external pion fields.
The leading one-body amplitude and subleading contact 
amplitude were obtained including the strong initial 
state interaction and the strong and Coulomb final-state 
interactions.  
After we fix the LECs for the $NN$ scatterings,
there remains only one unknown constant, $\tilde{d}_\pi^{(2)}$,  
in the amplitude. 
We estimated it from the one-pion exchange
diagram in the pionful theory. 
Although this method does not allow us
to fix the sign of $\tilde{d}_\pi^{(2)}$,
we have found that one of the two choices for $\tilde{d}_\pi^{(2)}$ 
leads to the cross sections that agree with
the experimental data reasonably well. 
On the other hand, the whole range 
of the experimental data near threshold
can be reproduced by adjusting the only
unknown LEC in the theory, $\tilde{d}_\pi^{(2)}$.
As discussed in Introduction, this is an expected result
because the energy dependence of the experimental total cross section
is known to be well described by the phase factor 
and the final-state interaction~\cite{metal-npa92},
which have been taken into account in this work,
and the overall strength of the cross section can be 
adjusted by the value of $\tilde{d}_\pi^{(2)}$.
This feature would be the same 
in the NNLO HB$\chi$PT calculations
because an unknown constant 
appears in the contact $\pi NNNN$ vertex 
and can be adjusted 
so as to reproduce the experimental data
(though there are many other corrections coming out
of the pion loop diagrams).

The main problem in the NNLO HB$\chi$PT calculations
that the contributions of formally different chiral orders 
become of comparable magnitude
has not been fully clarified in this work. 
Though the main difficulties in the HB$\chi$PT calculations 
may stem from the suppression mechanisms in the production operator,
another difficulty would stem from the heavy-baryon formalism
involving the typical large momentum transfer. 
The equations of motion of the heavy-field,
$v\cdot p\simeq \vec{p}^2/(2m_N)$,
makes a connection between 
the terms in the different orders, 
one derivative term $v\cdot p$ and 
two-derivative term $\vec{p}^2/(2m_N)$,
and thus the order counting rules become 
not transparent.\footnote{
When there is one heavy-particle, we can avoid the problem
by choosing the velocity vector $v^\mu$ so as to $v\cdot p =0$.
But if there are two heavy-particles, we cannot avoid
the problem.}
This is the reason why we needed to 
take account of the relativistic corrections:
e.g., 
the term proportional to $v\cdot \Delta \Delta \cdot D$ 
in ${\cal L}_{\pi N}^{(3)}$ in Eq.~(\ref{eq;L3})
is a relativistic correction to the term proportional to 
$(v\cdot \Delta)^2$ in ${\cal L}_{\pi N}^{(2)}$ in Eq.~(\ref{eq;L2}).
As mentioned in Introduction, one way 
to solve the problem
will be to employ the modified counting rules 
(see Refs.~\cite{letal-epja06,hw-plb07} as well)
and calculate the cross section 
collecting all of the pieces up to NNLO.
However, because a lot of higher order terms will be involved 
in the NNLO HB$\chi$PT calculation with the modified counting rules,
another possible way to calculate the production operator  
may be to employ a manifestly 
Lorentz invariant baryon chiral perturbation theory~\cite{af-prd07} 
with an additional subtraction scheme, 
such as the infrared renormalization scheme~\cite{bl-epjc99} or 
the extended on-mass shell scheme~\cite{fgjs-prd03}. 
That would be worth studying
in order to clarify the issues pertaining to the  HB$\chi$PT calculation
of the near-threshold $pp\to pp\pi^0$  reaction
up to one-loop order. 

\vskip 2mm \noindent
{\bf Acknowledgments}

The author would like to thank Ch. Elster for the discussion
which inspired this work,
T. Sato for discussion and communications,
and S.~X. Nakamura, C.~H. Hyun, F. Myhrer, K. Kubodera,
C. Hanhart, and U.-G. Mei\ss ner 
for carefully reading the manuscript and comments on it.
The author would like to thank the Institution for Nuclear
Theory at the University of Washington for its hospitality and 
the Department of Energy for partial support during the completion 
of this work.
This work is supported by Korean Research Foundation and The
Korean Federation of Science and Technology Societies Grant 
founded by Korean Government 
(MOEHRD, Basic Research Promotion Fund): the Brain Pool program
(052-1-6) and KRF-2006-311-C00271.

\end{document}